\documentclass[twocolumn,aps,prb,superscriptaddress]{revtex4-2}
\usepackage{amsfonts}
\usepackage{amsmath}
\usepackage{amssymb}
\usepackage{graphicx}
\usepackage{color}
\usepackage{ulem}
\usepackage[colorlinks, urlcolor=blue, citecolor=blue, linkcolor=magenta]{hyperref}

\begin{document}

\title{Quantum Mpemba effect in Local Gauge Symmetry Restoration}
\author{Hao-Yue Qi}
\email{qhaoyue@mail.ustc.edu.cn}
\affiliation{Hefei National Research Center for Physical Sciences at the Microscale and
School of Physical Sciences, University of Science and Technology of China,
Hefei 230026, China}
\affiliation{CAS Center for Excellence in Quantum Information and Quantum Physics,
University of Science and Technology of China, Hefei 230026, China}
\affiliation{Hefei National Laboratory, University of Science and Technology of China,
Hefei 230088, China}
\author{Wei Zheng}
\email{zw8796@ustc.edu.cn}
\affiliation{Hefei National Research Center for Physical Sciences at the Microscale and
School of Physical Sciences, University of Science and Technology of China,
Hefei 230026, China}
\affiliation{CAS Center for Excellence in Quantum Information and Quantum Physics,
University of Science and Technology of China, Hefei 230026, China}
\affiliation{Hefei National Laboratory, University of Science and Technology of China,
Hefei 230088, China}
\date{\today }

\begin{abstract}
Understanding relaxation in isolated quantum many-body systems remains a central challenge. Recently, the quantum Mpemba effect (QME)—a counterintuitive relaxation phenomenon—has attracted considerable attention and has been extensively studied in systems with global symmetries. 
Here, we study the QME in gauge theories with massive local gauge symmetries.
In the lattice Schwinger model, we demonstrate that the gauge structure of the reduced density matrix of a subsystem is entirely determined by the initial state and remain unchanged during the time evolution.
We then investigate whether gauge symmetry can be dynamically restored following a symmetric quench. 
Analytical and numerical results show that when the Maxwell term is zero, gauge symmetry restoration fails due to the emergence of a peculiar conservation law. However,
for any finite Maxwell term, subsystem gauge symmetry is restored in the thermodynamic limit.
Based on these results, we systematically construct a families of initial states exhibiting the QME. We further explore the QME in the quantum link model-a truncated lattice Schwinger model, which has been realized in experiments. Moreover, we propose an experimentally accessible order parameter that correctly captures the QME.
Our work demonstrates the generality of the quantum Mpemba effect even in the local gauge symmetries, and are directly relevant to ongoing quantum-simulation experiments of gauge theories.
\end{abstract}
\date{\today}
\maketitle

\section{Introduction}
Understanding the relaxation of isolated quantum many-body systems is a central issue in non-equilibrium statistical physics. 
In particular, while the eigenstate thermalization hypothesis (ETH) provides conditions for relaxation, the relevant timescales remain elusive \cite{Deutsch@1991,Srednicki@1999,Rigol@2008,Vengalattore@2011,Alessio@2016}.
In this context, diverse nonequilibrium phenomena have emerged, revealing the rich structure of relaxation dynamics. 
One striking example is the Mpemba effect—a counterintuitive phenomena originally observed in classical physics \cite{Mpemba@1969}. 
In isolated systems, the quantum Mpemba effect (QME) manifests as anomalous relaxation dynamics \cite{Calabrese@2023,Murciano@2025,Zhang@2025}: under a symmetric quench, a more asymmetric initial state restores subsystem symmetry faster than a more symmetric one. 
The QME has been observed in trapped-ion quantum simulators \cite{Joshi@2024} and extensively studied in both integrable \cite{Calabrese@2024(a),Calabrese@2024(b),Calabrese@2024(c),Bertini@2024,Ares@2025,Sengupta@2025} and chaotic systems \cite{Zhang@2024,Yao@2024,Luca@2025,Bertini@2025(a),Bertini@2025(b),Piroli@2025,Bhore@2025,Fan@2025} with Abelian and non-Abelian global symmetries. 
While global symmetries underpin the notable quantum Mpemba effect, their role in gauge theories—where extensive local constraints arise—remains largely unexplored.

Gauge theories play a pivotal role across physics, from particle interactions \cite{Weinberg@1995,Kogut@1979} to emergent phenomena in quantum materials \cite{Fradkin@2013}. Recent advances in quantum technologies—ranging from trapped ions \cite{Blatt@2016,Zoller@2019,Monroe@2021,Ringbauer@2025,Cetina@2025} and superconducting qubits \cite{Fan@2022,Hauke@2025,Rico@2025,Cochran@2025} to ultracold atoms in optical lattices \cite{Bloch@2019,Yang@2020,Mil@2020,Zhou@2022,Wang@2023,Yuan@2025,Liu@2025} and Rydberg atom arrays \cite{Lukin@2017,Dalmonte@2020,Zoller@2025}—now provide powerful platforms to simulate gauge theories far from equilibrium. Unlike conventional formulations in particle physics, gauge charges in these artificial platforms are not restricted to physical sector. They may fluctuate due to intersector superposition or unwanted gauge-violation noise. This freedom gives rise to novel effects \cite{Desaules@2023,Desaules2@2023,Ciavarella@2025,Xu@2025,Cataldi@2025,Grover@2016,Smith@2017,Heyl@2018,
Smith@2020,Zhai@2020,Chen@2023,Zheng@2024a,Zheng@2024b,Davoudi@2025}, including disorder-free localization, exotic phase transitions, and gauge-violation spectra. Such freedom also raises the question of whether a gauge-violation state, which breaks the gauge symmetry, can dynamically restore the local gauge symmetry during the relaxation, and whether the QME can emerge from such a restoration process. Such considerations are central to recent studies of the QME \cite{Calabrese@2023,Murciano@2025,Zhang@2025,Joshi@2024,Calabrese@2024(a),Calabrese@2024(b),
Calabrese@2024(c),Bertini@2024,Ares@2025,Sengupta@2025,Zhang@2024,Yao@2024,Luca@2025,Bertini@2025(a),Bertini@2025(b),
Piroli@2025,Bhore@2025,Fan@2025}
and resonate with ongoing quantum simulations of gauge theories \cite{Blatt@2016,Zoller@2019,Monroe@2021,Ringbauer@2025,Cetina@2025,Fan@2022,Hauke@2025,Rico@2025,
Cochran@2025,Bloch@2019,Yang@2020,Mil@2020,Zhou@2022,Wang@2023,Yuan@2025,Liu@2025, Lukin@2017,Dalmonte@2020,Zoller@2025}.

In this work, we investigate the local gauge-symmetry restoration and the associated QME in synthetic lattice gauge theory.
Using (1+1)-dimensional lattice Schwinger model as a prototypical example, we show that, owing to the locality of gauge symmetry, the symmetry structure of the reduced density matrix is time independent and is therefore entirely determined by the initial state. 
As a result, gauge theories provide a particularly suitable setting for studying the QME, in which gauge-symmetry restoration is more tractable than the restoration of global symmetries.
Combining analytical and numerical approaches, we demonstrate that subsystem gauge symmetry can be restored in the thermodynamic limit for any finite Maxwell term. 
By contrast, when the Maxwell term is zero, symmetry restoration is absent due to the emergence of an additional local symmetry, which renders the energy spectrum degenerate across different gauge sectors. 
In the regime where symmetry restoration occurs, we systematically construct families of initial states that exhibit the QME. 
Since the Hilbert space of the lattice Schwinger model is infinite dimensional—even for a single link—its direct experimental simulation is challenging. We therefore further explore the QME in the quantum link model, which is obtained by truncating the allowed eigenvalues of the electric field. 
In the above, we quantify gauge symmetry breaking using entanglement asymmetry, which is generally difficult to measure experimentally. Thus, we  propose an experimentally measurable order parameter in the quantum link model that qualitatively captures the dynamics of the QME.

The organization of this paper is as follows. In Sec.~\ref{II}, we introduce the lattice Schwinger model and the quench protocol considered in this work, and analyze the characteristic structure of the reduced density matrix in gauge theories. In Sec.~\ref{III}, we study gauge-symmetry restoration from a dephasing perspective and construct initial states exhibiting the QME. In Sec.~\ref{IV}, we investigate the QME in the experimentally realized quantum link model and propose an order parameter to detect the QME. Finally, we summarize our results in Sec.~\ref{V}.

\section{Model and setup}\label{II}
\begin{figure}[t]
  \centering
  \includegraphics[width=\columnwidth]{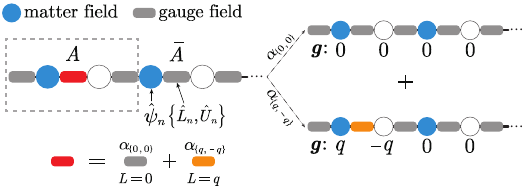}
  \caption{Initial state configuration of the lattice Schwinger model.
The matter field is initialized in a N\'{e}el state, consisting of alternating empty and occupied fermion sites. The gauge field on a single link within subsystem $A$ is prepared in a superposition of flux states with coefficients $\alpha_{\boldsymbol{g}_A}$, resulting in an overall initial state that is a superposition of distinct gauge sectors.}\label{fig1}
\end{figure}
\subsection{The lattice Schwinger model}
We consider the lattice Schwinger model with U(1) gauge symmetry \cite{Schwinger@1962a,Schwinger@1962b}. This model describes the (1+1)-dimensional lattice quantum electrodynamics, and displays many paradigmatic features also found in more complex gauge theories, such as confinement-deconfinement transition and string-breaking dynamics. The system is governed by the Kogut–Susskind Hamiltonian \cite{Kogut@1975},
\begin{equation}\label{SM}
\begin{aligned}
\hat{H}=&-w \sum_{n=1}^{N-1}\left(\hat{\psi}_n^{\dagger} \hat{U}_n \hat{\psi}_{n+1}+\text { h.c. }\right)\\
&+m \sum_{n=1}^N(-1)^n \hat{\psi}_n^{\dagger} \hat{\psi}_n+J \sum_{n=1}^{N-1}\left(\hat{L}_n+\frac{\theta}{2 \pi}\right)^2,
\end{aligned}
\end{equation}
where $\hat{\psi}_n$ is the fermionic annihilation operator at lattice site $n$, and $\hat{U}_n=e^{i\hat{\varphi}_n}$ is the $U(1)$ parallel transporter on bond $(n,n+1)$. The electric field operator is defined as
$\hat{L}_n=-i\partial/\partial\hat{\varphi}_n$ satisfying $[\hat{L}_n,\hat{U}_n]=\hat{U}_n$, which guarantees gauge invariance. The topological angle $\theta\in[0,2\pi]$ describes a constant classical background field. For $|\theta|<\pi$, the system is in confinement phase, whereas at $|\theta|=\pi$ it is in deconfinement phase.

The central concept in gauge theories is the presence of local gauge symmetries, which are encoded as a set of local constraints on the lattice. Specifically, the generators of these symmetries are given by the Gauss-law operators
\begin{equation}\label{GL}
  \hat{G}_n=\hat{L}_n-\hat{L}_{n-1}-\hat{\psi}_n^{\dagger} \hat{\psi}_n+\frac{1}{2}\left[1-(-1)^n\right],
\end{equation}
which commute with the Hamiltonian, $[\hat{G}_n,\hat{H}]=0$.
These symmetries partition the Hilbert space into superselection sectors, each labeled by a set of gauge-charges $\boldsymbol{g}=\{g_1,\cdots,g_N\}$, i.e., $\hat{G}_{n}|\Psi_{\boldsymbol{ g}}\rangle=g_{n}|\Psi_{\boldsymbol{g}}\rangle$. States belonging to different gauge sectors are orthogonal, $\langle\Psi_{\boldsymbol{g}}|\Psi_{\boldsymbol{g}^\prime}\rangle=\delta_{\boldsymbol{g}\boldsymbol{g}^{\prime}}$.
Since the eigenvalues of $\hat{\varphi}_n$ are defined modulo $2\pi$, $\hat{L}_n$ has integer eigenvalues, and so do the gauge charges $g_n\in\mathbb{Z}$. The standard Gauss law requires these local charges to
be zero, corresponding to the so called physical sector. In this work, however, we consider unconstrained gauge theories and focus on states that are distributed across different gauge sectors. Throughout this work we impose open boundary conditions and fix the boundary field $\hat{L}_0=\hat{L}_N=0$. Under these conditions, the gauge constraints render all gauge-field degrees of freedom redundant, allowing them to be sequentially integrated out.

\subsection{Reduced density matrixes and entanglement asymmetry in gauge theory}
We explore the quantum Mpemba effect in isolated quantum systems following a global quench. Since the time evolution is unitary, the system does not equilibrate globally to a stationary state. Instead, one must focus on a subsystem $A$, which is described by a reduced density matrix.
We first analyze the structure of reduced density matrices in gauge theories.
For a generic initial state $\rho_{\text{ini}}$, its time evolution can be decomposed as
\begin{equation}\label{DensityMatrix}
\rho(t) = \sum_{\boldsymbol{g},\boldsymbol{g}^{\prime}}\rho_{\boldsymbol{g},\boldsymbol{g}^\prime}(t),
\end{equation}
with $\rho_{\boldsymbol{g},\boldsymbol{g}^\prime}(t)=e^{-i\hat{H}_{\boldsymbol{g}}t}\hat{\Pi}_{\boldsymbol{g}}
\rho_{\text{ini}}
\hat{\Pi}_{\boldsymbol{g}^{\prime}}e^{i\hat{H}_{\boldsymbol{g}^{\prime}}t}$, where $\hat{\Pi}_{\boldsymbol{g}}$ is the projector onto sector $\boldsymbol{g}$, and $\hat{H}_{\boldsymbol{g}}=\hat{\Pi}_{\boldsymbol{g}}\hat{H}$ denotes the Hamiltonian restricted to that sector.
We divide the system into $A\cup\bar{A}$, with gauge charges well defined at each position within $A$. For the lattice Schwinger model, this corresponds to choosing a subsystem that contains the lattice sites and the adjacent links associated with each site in $A$, as illustrated in Fig.~\ref{fig1}. The reduced density matrix $\rho_A=\mathrm{Tr}_{\bar{A}}\rho$ then takes the form
\begin{equation}
\rho_A(t)=\sum_{\boldsymbol{g}_A \boldsymbol{g}_A^{\prime}} \sigma_{\boldsymbol{g}_A, \boldsymbol{g}_A^{\prime}}(t)\label{RDM}.
\end{equation}
Here, we have define $\sigma_{\boldsymbol{g}_A, \boldsymbol{g}_A^{\prime}}(t)=\sum_{\boldsymbol{g}_{\bar{A}}\boldsymbol{g}_{\bar{A}}^{\prime}}
\mathrm{Tr}_{\bar{A}}[
\rho_{\boldsymbol{g},\boldsymbol{g}^{\prime}}(t)]$ and $\boldsymbol{g}=\{\boldsymbol{g}_A, \boldsymbol{g}_{\bar{A}}\}$. Owing to the locality of gauge symmetries, the partial trace over $\bar{A}$ does not alter the gauge charges within subsystem $A$ of full state.
Consequently, using the orthogonality of states belonging to different gauge sectors, $\sigma_{\boldsymbol{g}_A, \boldsymbol{g}_A^{\prime}}(t)$ only couples sectors with gauge charges $\boldsymbol{g}_A$ and $\boldsymbol{g}_A^{\prime}$ at arbitrary time $t$, satisfying $\hat{\Pi}_{\tilde{\boldsymbol{g}}_{A}}\sigma_{\boldsymbol{g}_A, \boldsymbol{g}_A^{\prime}}(t)\hat{\Pi}_{\tilde{\boldsymbol{g}}^{\prime}_A}=\delta_{\tilde{\boldsymbol{g}}_{A},\boldsymbol{g}_A}
\delta_{\tilde{\boldsymbol{g}}^{\prime}_{A},\boldsymbol{g}_A^{\prime}}\sigma_{\boldsymbol{g}_A, \boldsymbol{g}_A^{\prime}}(t)$. Thus, the gauge-sector structure of $\rho_A(t)$ is entirely determined by the full density matrix $\rho(t)$, and no new sectors are generated during time evolution. This implies that, in gauge theories one can directly engineer which sector appear in the time evolution of $\rho_A(t)$ by preparing initial state $\rho_{\text{ini}}$ with or without the associated components. In contrast, for global symmetries where the conserved charge is delocalized, tracing out $\bar{A}$ generally mixes charge sectors, and the structure of $\rho_A(t)$ evolves with time. The above analysis is generic and does not rely on model-specific details. The sample structure makes the investigation of symmetry restoration and the QME in the following more clear than global ones.

In general, $\rho_A(t)$ breaks the gauge symmetry, i.e., $[\rho_A(t), \hat{G}_{n}]\neq0$ for some $n\in A$. We employ the entanglement asymmetry (EA) \cite{Calabrese@2023} to quantify the degree of symmetry breaking in subsystem $A$,
\begin{equation}\label{EA}
  \Delta S_A(t)=S[\rho_{A,G}(t)]-S[\rho_A(t)],
\end{equation}
where $S[\rho_A]=-\mathrm{Tr}(\rho_A\log\rho_A)$ denotes the von-Neumann entropy of subsystem $\rho_A$,
and $\rho_{A,G}(t)=\sum_{\boldsymbol{g}_A}\sigma_{\boldsymbol{g}_A, \boldsymbol{g}_A}(t)$ is the symmetric counterpart of $\rho_A(t)$ obtained by projecting onto gauge sectors $\boldsymbol{g}_A$. By construction, $\Delta S_A(t)\geq0$, and it vanishes only when the gauge symmetry is already restored, $\rho_A(t)=\rho_{A,G}(t)$. The QME occurs when an initial state with a larger EA restores subsystem symmetry faster than another state with a smaller EA. We further confirm in the Appendix that other metrics quantifying subsystem symmetry breaking produce the same relaxation behavior.

In the following, we consider a family of initial states of the form
\begin{equation}\label{InitialState}
  |\Psi_\text{ini}(\boldsymbol{q}_A)\rangle=\alpha_{\boldsymbol{0}_{A}} |\Psi_{\boldsymbol{0}_A}\rangle
  +\alpha_{\boldsymbol{q}_{A}} |\Psi_{\boldsymbol{q}_A}\rangle.
\end{equation}
For each sector, we take $|\Psi_{\boldsymbol{g}_A}\rangle=
|\phi^{\text{N\'{e}el}}\rangle|L_{\{\boldsymbol{g}_A,\boldsymbol{0}_{\bar{A}}\}}\rangle$,
where the fermionic field is prepared in the N\'{e}el state $|\phi^{\text{N\'{e}el}}\rangle=|1010\cdots\rangle$, representing the bare staggered vacuum. The gauge field configuration $|L_{\{\boldsymbol{g}_A,\boldsymbol{0}_{\bar{A}}\}}\rangle$ is uniquely fixed by gauge charges
$\{\boldsymbol{g}_A,\boldsymbol{0}_{\bar{A}}\}$.
Here, complement $\bar{A}$ carries vanishing gauge charges, while $A$ is prepared in a superposition of different gauge sectors, explicitly breaking gauge symmetry in $A$. A schematic illustration of the initial state is shown in Fig.\ref{fig1}.

\section{Gauge symmetry restoration and quantum Mpemba effect}\label{III}
\begin{figure}[t]
  \centering
  \includegraphics[width=\columnwidth]{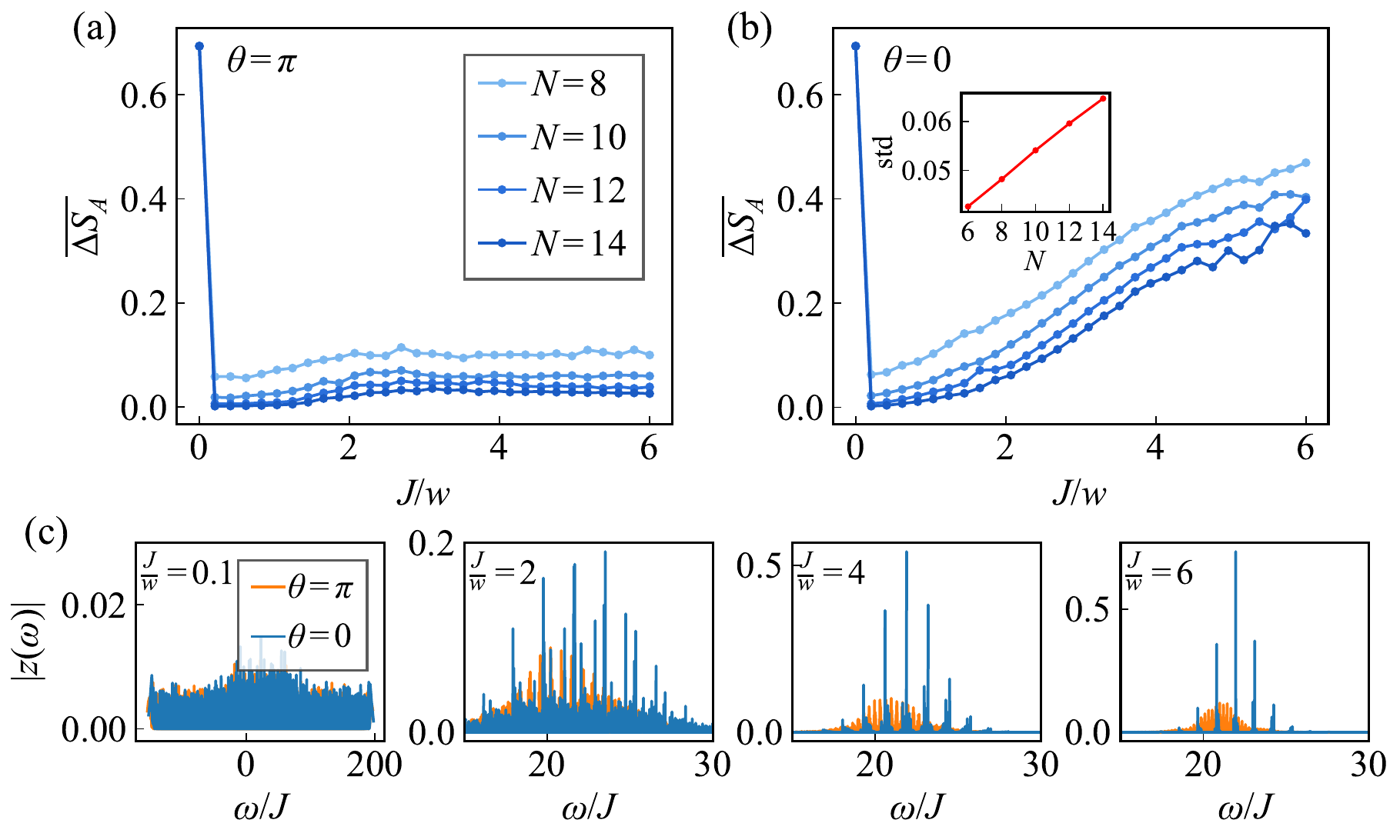}
  \caption{The long-time equilibrium values of the EA, averaged over $wt=10^3-10^4$, as a function of the coupling strength $J$ at topological angle $\theta=\pi$ (a) and $0$ (b) for different system size $N$. (c) The distribution $|z(\omega)|$. At $\theta=\pi$, $|z(\omega)|$ remains densely distributed for all $J$, whereas at $\theta=0$, it collapses into discrete peaks equally spaced by $J$ at large couplings.
  The inset in (b) shows the standard deviation of a single peak increasing with the system size at fixed value of $J/w=6$. }\label{fig2}
\end{figure}
Since the locality of gauge symmetries, it is not a priori clear whether this local symmetry can be restored at the subsystem level. We therefore first investigate whether gauge symmetry can be restored by a symmetric quench. Having identified the parameter regimes in which symmetry restoration occurs, we then construct families of initial states that exhibit the quantum Mpemba effect.
To this end, we consider a subsystem consisting of two sites together with the adjacent links on the left side of the system. The initial state in Eq. \eqref{InitialState} is chosen as an equal superposition of two gauge sectors, with $\alpha_{\boldsymbol{0}_A}=\alpha_{\boldsymbol{q}_A}=1/\sqrt{2}$ and $\boldsymbol{q}_A=\{1,-1\}$.
Focusing on vanishing fermion mass $m=0$, we analyze the long-time equilibrium values of the EA as a function of the coupling $J$.
The numerical results are shown in Fig.~\ref{fig2}(a-b). For zero Maxwell term $J=0$, the equilibrium values $\overline{\Delta S_A}$ remains finite for both $\theta=0$ and $\theta=\pi$, indicating the absence of gauge-symmetry restoration. For finite $J$, the values crucially depend on the topological angle.
At $\theta=\pi$, $\overline{\Delta S_A}$ decays to zero with increasing system size for small $J$, and remains small finite value for large $J$. In contrast, at $\theta=0$ it approaches zero for small $J$, but remains appreciably finite at larger $J$ within the accessible system sizes. Below, we provide a general theoretical analysis that accounts for these observations.

\subsection{Theoretical analysis of gauge-symmetry restoration}
Gauge symmetry restoration means that the off-diagonal blocks $\sigma_{\boldsymbol{g}_A, \boldsymbol{g}_A^{\prime}}(t)$ with $\boldsymbol{g}_A\ne\boldsymbol{g}_A^{\prime}$ in Eq.~\eqref{RDM} decay to zero at long times. Equivalently, for any operator $\hat O_A$ in subsystem $A$, their Hilbert–Schmidt inner product $\mathrm{Tr}_{A}[\hat{O}_A\sigma_{\boldsymbol{g}_A, \boldsymbol{g}_A^{\prime}}\left(t\right)]$ vanishes asymptotically.
Expanding this quantity in the orthonormal eigenbasis ${|E_{\boldsymbol{g}n}\rangle}$ of the Hamiltonian, one obtains
\begin{equation}
\begin{aligned}
  \mathrm{Tr}_{A}[\hat{O}_A\sigma_{\boldsymbol{g}_A, \boldsymbol{g}_A^{\prime}}(t)]
  &= \sum_\alpha v_{\alpha} e^{-i \omega_\alpha t} \\
  &= \sum_{\omega} z(\omega) e^{-i \omega t}.
\end{aligned}
\end{equation}
Here, the energy gaps are $\omega_\alpha=E_{\boldsymbol{g}n}
-E_{\boldsymbol{g}^\prime n^\prime}$, where $E_{\boldsymbol{g}n}$ denotes the $n$th eigenenergy in sector $\boldsymbol{g}$, and $\alpha=(n,n^\prime,\boldsymbol{g}_{\bar{A}},\boldsymbol{g}_{\bar{A}}^{\prime})$ labels all possible gaps. The coefficients read $v_\alpha=\langle E_{\boldsymbol{g}n}|\rho_{\text{ini}} |E_{\boldsymbol{g}^{\prime}n^{\prime}}\rangle\langle E_{\boldsymbol{g}^{\prime}n^{\prime}}|\hat{O}_A |E_{\boldsymbol{g}n}\rangle$. In the second equation, we grouped them as $z(\omega)=\sum_{\alpha}v_{\alpha}\delta_{\omega_{\alpha},\omega}$. Thus, the restoration dynamics can be viewed as a superposition of complex vectors $z(\omega)e^{-i\omega t}$ rotating with frequency $\omega$. Gauge symmetry is restored when these vectors dephase and sum to zero.
Because the gauge-sector structure of the reduced density matrix is time independent, the dynamics involve only off-diagonal block $\sigma_{\boldsymbol{0}_A,\boldsymbol{q}_A}(t)$ for the initial state Eq.\eqref{InitialState}.

All dynamical information is encoded in the distribution $z(\omega)$.
Gauge symmetry restoration requires the zero-frequency component $z(0)$ vanishes in the thermodynamic limit. At $J=0$, the gauge field has no dynamics and acts as a static background. In our model, this makes the system acquires an additional local symmetry, $[\hat{H},\hat{U}_n]=0$ for all $n$. Thus, if $|E_{\boldsymbol{g}n}\rangle$ is an eigenstate with energy $E_{\boldsymbol{g}n}$ in sector
$\boldsymbol{g}_A=\{1,-1\}$, then $\hat U_1|E_{\boldsymbol{g}n}\rangle$ is also an eigenstate with the same energy but in sector $\boldsymbol{g}_A'=\{0,0\}$. This exact degeneracy between different gauge sectors prevents dephasing and leads to a finite zero-frequency contribution $z(0)$. More explicitly, expanding the eigenstates in an orthonormal basis,
$|E_{\boldsymbol{g}n}\rangle=\sum_{i}\beta_{n,i}|\phi_i\rangle|L_{\boldsymbol{g}i}\rangle$, the coefficients $\beta_{n,i}$ are independent of $\boldsymbol{g}$ because $\hat U_1$ acts only on the gauge-field degrees of freedom. Since $\hat {O}_A$ is an arbitrary operator acting on $A$, we may take
$\hat{O}_A=\sum_{ii^\prime}c_{ii^\prime}|L^A_{\boldsymbol{0}_Ai^{\prime}}\rangle|\phi^A_{i^\prime}\rangle
\langle\phi^A_{i}|\langle L^A_{\boldsymbol{q}_Ai}|$ with arbitrary coefficients $c_{ii^\prime}$. Then, substituting the initial state in Eq.~\eqref{InitialState} into the expression for $z(\omega)$, zero-frequency component $z(0)$ can be derived explicitly as
\begin{equation}
  |z(0)|=|\alpha_{\boldsymbol{q}_A}\alpha_{\boldsymbol{0}_A}^{*}
\mathrm{Tr}(\rho_D\hat{O}_A^f)|>0,
\end{equation}
where $\rho_D=\sum_{n}|c_n|^2|E^f_n\rangle\!\langle E^f_n|$ is the diagonal ensemble with $c_n=\langle E^f_n|\phi^{\text{N\'{e}el}}\rangle$. Here, we have define the fermionic part of eigenstate as
$|E^f_n\rangle=\sum_{i}\beta_{n,i}|\phi_i\rangle$, and fermionic operator as $\hat{O}^f_A=\sum_{ii^\prime}c_{ii^\prime}|\phi^A_{i^\prime}\rangle
\langle\phi^A_{i}|$. This demonstrates the absence of gauge-symmetry restoration at $J=0$ observed in Fig.~\ref{fig2}(a-b). The same conclusion holds for any initial state with identical fermionic field configurations across sectors, as shown in the Appendix.

At $J\ne0$, we numerically calculate the distribution $z(\omega)$, shown in Fig.~\ref{fig2}(c). Since $\hat{O}_A$ is an arbitrary local operator, we have assumed $\langle E_{(\boldsymbol{g}_A^\prime,\boldsymbol{g}_{\bar{A}})n^\prime}|\hat{O}_A
|E_{(\boldsymbol{g}_A,\boldsymbol{g}_{\bar{A}})n}\rangle\sim O(1)$. At $\theta=\pi$, $|z(\omega)|$ remains densely distributed within a finite frequency window for all $J$, enabling effectively dephasing and hence leading to small EA. By contrast, at $\theta=0$, $|z(\omega)|$ is dense only at small $J$, while for large $J$ it collapses into discrete peaks equally spaced by $J$, suppressing dephasing and preventing restoration.
This contrasting behavior can be understood from perturbation theory. At $m=w=0$, the eigenstates of the Schwinger model Eq.~\eqref{SM} are product states of gauge fields, producing highly degenerate manifolds and hence discrete frequency peaks. Turning on $w$ lifts these degeneracies and broadens the peaks. At $\theta=\pi$, the perturbation acts already at first order, since the manifolds contain resonant states, such as $|0,-1,0,\cdots\rangle$ and $|0,0,0,\cdots\rangle$ which share the same energy $J(N-1)/4$. By contrast, at $\theta=0$ no such resonances exist and the first-order contribution vanishes, so the peaks remain much sharper at small $w/J$. However, since the peak also broadens with system size due to the growing manifold dimension, gauge symmetry is expected to be ultimately restored at any finite $J$ in the thermodynamic limit. Alternatively, since we probe subsystem dynamics, the equilibrium reduced density matrix behaves as a (generalized) Gibbs Ensemble. Therefore, a intuitive explanation for gauge symmetry restoration is the Elitzur's theorem—the impossibility of spontaneously breaking a local symmetry. Our results demonstrate concretely the gauge symmetry can be restored except at $J=0$ where a peculiar conservation laws emerges, consistent with Fig.~\ref{fig2}(a-b).

\subsection{Engineering quantum Mpemba effect}
\begin{figure}[t]
  \centering
  \includegraphics[width=\columnwidth]{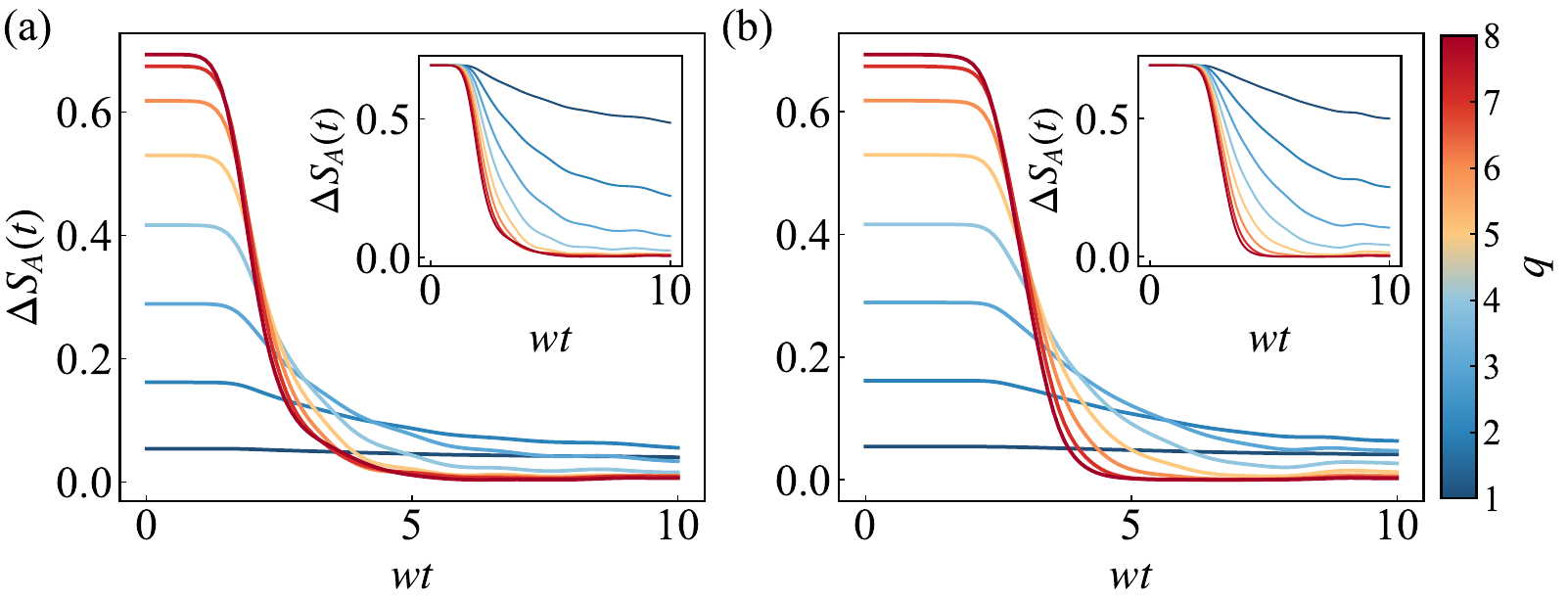}
  \caption{(a) and (b) show the EA dynamics for a two-site subsystem with $J=0.15$ and a three-site subsystem with $J=0.05$ in a system of size $N=16$. The insets display dynamics of initial states prepared as equal superpositions of gauge sectors, where larger $q$ leads to faster decay. By tuning the superposition coefficients, the QME is realized.}\label{fig3}
\end{figure}
Previous studies of the QME lacked a systematic way to identify or design such states. Here, in gauge theory we systematically construct families of initial states exhibiting the QME. Specifically, we consider a set of initial states in Eq.~\eqref{InitialState} for two-site subsystem $A$, $|\Psi_{\text{ini}}(q)\rangle=\sum_{g_A\in\{0,q\}}
\alpha_{\{g_A,-g_A\}}|\Psi_{\{g_A,-g_A\}}\rangle$ with $q=1,2,\cdots,q_{\text{max}}$. From Eq.~\eqref{RDM}, the corresponding reduced density matrix takes a $2\times2$ block structure spanned by the gauge sectors $\{0,0\}$ and $\{q,-q\}$.
For equal superpositions $\alpha_{\{0,0\}}=\alpha_{\{q,-q\}}$, the initial reduced density matrix reads
$\rho_A(0)=\frac{1}{2}(\mathbb{I}+\sigma^x)\otimes|\phi^{\text{N\'{e}el}}_A\rangle\langle\phi^{\text{N\'{e}el}}_A|$, which is independent of $q$.
Thus, the EA is identical for all $|\Psi_{\text{ini}}(q)\rangle$, taking value $\Delta S_A(0)=\log 2$.
More generally, for initial states with identical physical (fermionic) configurations across gauge sectors and vanishing gauge charges in $\bar{A}$, the EA depends only on the number of gauge sectors present in $\rho_A$
(see Appendix). Consequently, it suffices to prepare initial states with the same number of gauge sectors to ensure identical initial EA.
The EA dynamics is shown in the inset of Fig.~\ref{fig3}(a).
Crucially, we find that the rate of gauge symmetry restoration increases with the gauge charge $q$. Therefore we can always choose the coefficients as $\alpha_{\{0,0\}}=\cos\theta$ and $\alpha_{\{q,-q\}}=\sin\theta$ with $\theta=\tfrac{\pi}{4}\tfrac{q}{q_{\text{max}}}$, such that states with faster restoration are initialized with stronger symmetry breaking. This tuning naturally gives rise to the Mpemba effect shown in Fig.~\ref{fig3}(a). Moreover, this construction can straightforwardly generalizes to other families of initial states, such as those involving a three-site subsystem, $|\Psi_{\text{ini}}\rangle=\sum_{g_A\in\{0,q\}}
\alpha_{\{g_A,0_A,-g_A\}}|\Psi_{\{g_A,0_A,-g_A,\boldsymbol{0}_{\bar{A}}\}}\rangle$, as illustrated in Fig.~\ref{fig3}(b).

\section{Experimental consideration}\label{IV}
In the lattice Schwinger model, the Wilson's parallel transporter $\hat{U}_n\in U(1)$ is a continuous classical variable, satisfying $[\hat{U}_n,\hat{U}_n^\dag]=0$. As a consequence, the Hilbert space of this gauge theory is infinite-dimensional, even on a single link. From the perspective of experimental realizations, however, it is desirable to work with finite-dimensional local Hilbert spaces. This can be achieved by truncating the allowed eigenvalues of the electric field operator $\hat{L}_n$. Here, we restrict $\hat{L}_n$ to take values $0$ and $\pm1$.
This truncation implies the following replacement: $\hat{U}_n\rightarrow \hat{S}_n^+$ and $\hat{L}_n\rightarrow \hat{S}_n^z$, where $\hat{S}_n^+$ and $\hat{S}_n^z$ denote the spin-$1$ raising operator and the
$z$-component of the spin respectively. Moreover, we perform a particle hole transformation at odd lattice sites and a $\pi$ rotation around the $y$ axis on the gauge fields at even sites,
\begin{equation}
\begin{aligned}
n \ \text{odd}: \quad & \hat{\psi}_n \leftrightarrow  \hat{\psi}_n^{\dagger}, \\
n \ \text{even}: \quad &
\hat{S}_{n,n+1}^z \mapsto -\hat{S}_{n,n+1}^z,
\hat{S}_{n,n+1}^{\pm} \mapsto -\hat{S}_{n,n+1}^{\mp}.
\end{aligned}
\end{equation}
These transformations yield the quantum link model,
\begin{equation}\label{QLM}
\begin{aligned}
  H=-&w\sum_{n}\left(\psi_n S^+_n\psi_{n+1}+\mathrm{h.c.}\right)\\
  +&m\sum_{n}\psi_{n}^\dag\psi_n+J\sum_{n}((-1)^{n+1}S^z_n+\frac{\theta}{2\pi})^2.
\end{aligned}
\end{equation}
Despite the truncation of the electric field, this quantum link formulation preserves an exact local
U(1) gauge symmetry. The Gauss-law operator now becomes
\begin{equation}
\hat{G}_n=(-1)^{n+1}(\hat{S}^z_n + \hat{S}^z_{n-1}+\hat{\psi}_n^{\dagger} \hat{\psi}_n).
\end{equation}
The quantum link model has been experimentally realized on quantum simulators based on trapped ions \cite{Blatt@2016} or ultracold bosons \cite{Yang@2020,Mil@2020,Zhou@2022}. Many striking phenomena have been observed, including quark confinement \cite{Yuan@2025} and string-breaking dynamics \cite{Liu@2025}.

\begin{figure}
  \centering
  \includegraphics[width=\columnwidth]{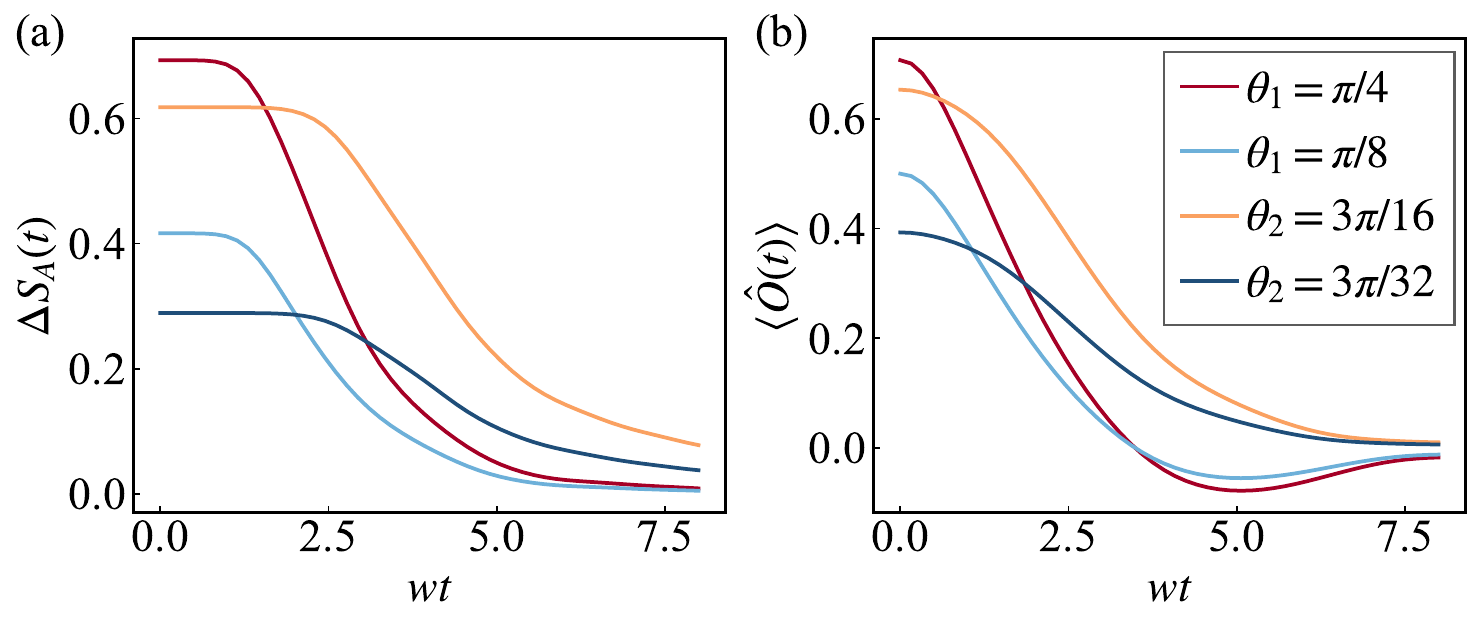}
  \caption{Panels (a) and (b) display the quantum Mpemba effect in the quantum link model with $N=14, J/w=0.2$ and $m=0$. In panel (a), gauge-symmetry breaking is quantified using the entanglement asymmetry, while in panel (b) it is characterized by the order parameter. Both diagnostics yield the same qualitative behavior.}\label{fig4}
\end{figure}

In this model, we can also construct families of initial states that exhibit the quantum Mpemba effect following the procedure outlined above. Specifically, we consider the following initial states involving different gauge sectors,
\begin{equation}\label{iniQLM}
\begin{aligned}
|\Psi_1(\theta_1\rangle=&\cos \theta_1|\Psi_{\{1,-1\}}\rangle+\sin \theta_1|\Psi_{\{0,0\}}\rangle, \\
|\Psi_2(\theta_2)\rangle=&\cos \theta_2|\Psi_{\{1,-1\}}\rangle+\sin \theta_2|\Psi_{\{-1,1\}}\rangle.
\end{aligned}
\end{equation}
For each sector, we take $|\Psi_{\{\boldsymbol{g}_A\}}\rangle=|00 \cdots 0\rangle|\boldsymbol{S}_{\{\boldsymbol{g}_A,\boldsymbol{0}_{\bar{A}}\}}\rangle$, where the fermionic field is prepared in the bare vacuum. The gauge field configuration $|\boldsymbol{S}_{\{\boldsymbol{g}_A\}}\rangle$ is
uniquely fixed by gauge charges $\{\boldsymbol{g}_A,\boldsymbol{0}_{\bar{A}}\}$. For $\theta_1=\theta_2$, the two initial states possess the same entanglement asymmetry but different relaxation rates.  By tuning the parameters such that $\theta_1<\theta_2<\pi/4$, we obtain the QME as shown in Fig.~\ref{fig4}(a).

In the above, we quantify the distance between two states by the EA. However, its experimental detection typically requires full state tomography, which remains extremely challenging in many-body systems. While randomized measurements are effective for detecting R\'{e}nyi etropy, they still demand precise control over individual qubits, which is difficult to achieve in many experimental platforms. Here, we instead propose an experimentally accessible order parameter,
\begin{equation}\label{OR}
  \hat{O}=\sum_{i\in A}\left[\hat{S}^x_i+\frac{(\hat{S^x_i})^2-(\hat{S}^y_i)^2}{\sqrt{2}}\right],
\end{equation}
which can be equivalently rewritten as $\hat{O}=\frac{1}{2}\sum_{i\in A}\left(S^+_i+\frac{S^+_iS^+_i}{\sqrt{2}}+\text{h.c.}\right)$. A nonvanishing expectation value of $\hat{O}$ thus provides a sufficient but not necessary condition for gauge symmetry breaking: If $\langle\hat{O}\rangle\ne0$, it signals the symmetry breaking, but the inverse is not true, which is a generic feature of order parameters used to characterize symmetry breaking. As shown in Fig.~\ref{fig4}(b), this order parameter correctly captures the crossing behavior characteristic of the QME, even though the precise crossing times differ from those obtained using the EA.

\section{Conclusions}\label{V}
In summary, we investigate whether gauge symmetry can be dynamically restored and whether the corresponding quantum Mpemba effect can emerge in gauge theories. Through numerical simulations and perturbative analysis of the lattice Schwinger model, we show that subsystem gauge symmetry is restored in the thermodynamic limit for any finite Maxwell term $J>0$, except at $J=0$ where an emergent local conservation law prevents restoration. Building on this result, we systematically construct families of initial states that exhibit the quantum Mpemba effect. From an experimental perspective, we further explore the quantum Mpemba effect in the quantum link model and propose an experimentally accessible order parameter that faithfully captures the Mpemba dynamics. Looking forward, our results indicate that the Mpemba effect hinges on distinct relaxation rates associated with different gauge sectors, which numerically appear to be positively correlated with the gauge-charge difference $|g_A-g_A^{\prime}|$, yet a rigorous demonstration for this dependence remains an open challenge.

\section{Acknowledgments}We thank Hui Sun for helpful discussion. This work
is supported by NSFC (Grants No. GG2030007011 and No.
GG2030040453) and Innovation Program for Quantum Science and Technology (and No.2021ZD0302004).


\newpage
\begin{widetext}
\appendix
\renewcommand{\theequation}{S\arabic{equation}}
\setcounter{equation}{0}
\section*{Appendix I: Derivation of the Reduced Dynamics}
\textbf{Gauge-symmetry restoration.}
Restoration of gauge symmetry corresponds to the decay of the
entanglement asymmetry (defined in the main text) to zero.
Equivalently, the off-diagonal blocks
$\sigma_{\boldsymbol{g}_A,\boldsymbol{g}_A^{\prime}}(t)$ with
$\boldsymbol{g}_A \neq \boldsymbol{g}_A^{\prime}$ must vanish at long
times.
To analyze this process we study their Hilbert–Schmidt inner product with an arbitrary operator,
\begin{equation}
\mathrm{Tr}_{A}\!\left[\,\hat{O}_A\,
\sigma_{\boldsymbol{g}_A,\boldsymbol{g}_A^{\prime}}(t)\right]
= \sum_{\boldsymbol{g}_{\bar{A}},\boldsymbol{g}_{\bar{A}}^{\prime}}
  \mathrm{Tr}\!\left[
  e^{-i\hat{H}_{\boldsymbol{g}}t}
  \hat{\Pi}_{\boldsymbol{g}}
  \rho_{\text{ini}}
  \hat{\Pi}_{\boldsymbol{g}^{\prime}}
  e^{i\hat{H}_{\boldsymbol{g}^{\prime}}t}
  \hat{O}_A \right].
\end{equation}
Expanding the projectors in the eigenbasis of each sector Hamiltonian,
$\hat{\Pi}_{\boldsymbol{g}}
 = \sum_{n}|E_{\boldsymbol{g}n}\rangle\langle E_{\boldsymbol{g}n}|$,
yields
\begin{equation}
\begin{aligned}
\mathrm{Tr}_{A}\!\left[\,\hat{O}_A\,
\sigma_{\boldsymbol{g}_A,\boldsymbol{g}_A^{\prime}}(t)\right]
&= \sum_{\boldsymbol{g}_{\bar{A}},\boldsymbol{g}_{\bar{A}}^{\prime},n,n^{\prime}}\!
   \langle E_{\boldsymbol{g}n}|
   \rho_{\text{ini}}
   |E_{\boldsymbol{g}^{\prime}n^{\prime}}\rangle\,
   \langle E_{\boldsymbol{g}^{\prime}n^{\prime}}|
   \hat{O}_A
   |E_{\boldsymbol{g}n}\rangle \,
   e^{-i\left(E_{\boldsymbol{g}n}
   -E_{\boldsymbol{g}^\prime n^\prime}\right)t} \\[6pt]
&= \sum_\alpha v_\alpha\,e^{-i\omega_\alpha t}
   = \sum_{\omega} z(\omega)\,e^{-i\omega t}.
\end{aligned}
\end{equation}
Here $\omega_\alpha = E_{\boldsymbol{g}n}
- E_{\boldsymbol{g}^\prime n^\prime}$ denotes the energy gaps, with
$\alpha=(n,n^\prime,\boldsymbol{g}_{\bar{A}},\boldsymbol{g}_{\bar{A}}^{\prime})$
labeling all possible contributions.
In the second line, terms with identical frequency are grouped as
$z(\omega)=\sum_{\alpha}v_{\alpha}\delta_{\omega_{\alpha},\omega}$.
Thus, the restoration dynamics can be understood as the interference of
complex vectors $z(\omega)e^{-i\omega t}$ rotating at frequency
$\omega$. Gauge symmetry is restored once these vectors dephase and
their sum approaches zero.

\textbf{Zero-frequency contribution.}
We consider an initial state with the same fermionic configuration in each sector:
\begin{equation}\label{SMini}
  |\Psi_{\text{ini}}\rangle=\sum_{\boldsymbol{g}}\alpha_{\boldsymbol{g}}|\Psi_{\boldsymbol{g}}\rangle
  =\sum_{\boldsymbol{g}i}\alpha_{\boldsymbol{g}}\beta_{i}|\phi_i\rangle|L_{\boldsymbol{g}i}\rangle,
\end{equation}
where $\beta_i$ is independent of $\boldsymbol{g}$. In the second equality, we expanded $|\Psi_{\boldsymbol{g}}\rangle = \sum_i \beta_i |\phi_i\rangle |L_{\boldsymbol{g} i}\rangle$, with the index $i$ labeling the fermionic product states $|\phi_i\rangle$. For the lattice Schwinger model, the gauge-field product state $|L_{\boldsymbol{g} i}\rangle$ is uniquely determined by the gauge charges $\boldsymbol{g}$ and the fermionic state $i$.
At $J=0$, all sector Hamiltonians $\hat{H}_{\boldsymbol{g}}$ share the same spectrum and fermionic configurations, i.e.,
\begin{equation}
  |E_{\boldsymbol{g}n}\rangle=\sum_{i}\beta_{n,i}|\phi_i\rangle|L_{\boldsymbol{g}i}\rangle,
\end{equation}
where $\beta_{n,i}$ is independent of $\boldsymbol{g}$. $E_{\boldsymbol{g}n}$ is the eigenenergy in sector $\boldsymbol{g}$ and satisfies $E_{\boldsymbol{g}n}=E_{\boldsymbol{g}^{\prime}n}$.

Assuming non-degenerate levels in each sector, the zero-frequency
contribution arises only from $n=n^\prime$.
Substituting the initial state Eq.~\eqref{SMini} into the expression of $z(0)$ gives
\begin{gather}
z(0)=
\sum_{\boldsymbol{g}_{\bar{A}},\boldsymbol{g}_{\bar{A}}^{\prime},n}\!
\alpha_{\boldsymbol{g}}\alpha_{\boldsymbol{g}^{\prime}}^{*}\langle E_{\boldsymbol{g}n}|\Psi_{\boldsymbol{g}}\rangle\langle\Psi_{\boldsymbol{g}^{}\prime}
|E_{\boldsymbol{g}^{\prime}n}\rangle\,\langle E_{\boldsymbol{g}^{\prime}n}|\hat{O}_A|E_{\boldsymbol{g}n}\rangle\notag\\
\langle E_{\boldsymbol{g}n}|\Psi_{\boldsymbol{g}}\rangle\langle\Psi_{\boldsymbol{g}^{}\prime}
|E_{\boldsymbol{g}^{\prime}n}\rangle=\sum_{ii^\prime}\beta_{n,i}^*\beta_i\beta_{i^\prime}^*
\beta_{n,i^\prime}\\
\langle E_{\boldsymbol{g}^{\prime}n}|\hat{O}_A|E_{\boldsymbol{g}n}\rangle=
\sum_{ii^\prime}
\beta_{n,i^{\prime}}^*\beta_{n,i}\langle\phi_{i^\prime}^{\bar{A}}|\phi_{i}^{\bar{A}}\rangle
\langle L_{\boldsymbol{g}_{\bar{A}}^{\prime} i^\prime}^{\bar{A}}|L_{\boldsymbol{g}_{\bar{A}} i}^{\bar{A}}\rangle
\langle L^A_{\boldsymbol{g}_A^{\prime}i^{\prime}}|\langle\phi^A_{i^\prime}|\hat{O}_A
|\phi^A_{i}\rangle| L^A_{\boldsymbol{g}_Ai^{\prime}}\rangle\notag
\end{gather}
Here, we introduce the fermionic part of initai state and eigenstate as $|\Psi^f_{\text{ini}}\rangle=\sum_{i}\beta_{i}
|\phi_i\rangle$ and $|E_n^f\rangle=\sum_{i}\beta_{n,i}|\phi_i\rangle$. Since $\hat{O}_A$ is an arbitrary operator acting within $A$, we may set
$\hat{O}_A=\sum_{ii^\prime}c_{ii^\prime}|L^A_{\boldsymbol{g}^{\prime}_Ai^{\prime}}\rangle|\phi^A_{i^\prime}\rangle
\langle\phi^A_{i}|\langle L^A_{\boldsymbol{g}_Ai}|$. Finally, we find
\begin{equation}
\begin{aligned}
  z(0)=&\sum_{\boldsymbol{g}_{\bar{A}},\boldsymbol{g}_{\bar{A}}^{\prime}}\mkern-8mu{}'
  \alpha_{\boldsymbol{g}}\alpha_{\boldsymbol{g}^{\prime}}^{*}
\sum_{n}\langle E_n^f|\Psi^f_{\text{ini}}\rangle\langle\Psi^f_{\text{ini}}|E_n^f\rangle\langle E_n^f|\hat{O}_A^f|E_n^f\rangle\\
=&C_{\boldsymbol{g}_{A},\boldsymbol{g}_{A}^{\prime}}
\mathrm{Tr}(\rho_D\hat{O}^f_A),
\end{aligned}
\end{equation}
where $\rho_D=\sum_{n}|c_n|^2|E^f_n\rangle\!\langle E^f_n|$ is the diagonal ensemble with $c_n=\langle E^f_n|\Psi^f_{\text{ini}}\rangle$, and $\hat{O}^f_A=\sum_{ii^\prime}c_{ii^\prime}|\phi^A_{i^\prime}\rangle
\langle\phi^A_{i}|$. Here, we have used the fact
$\langle \phi^{\bar{A}}_i | \phi^{\bar{A}}_{i'} \rangle \langle L^{\bar{A}}_{\boldsymbol{g}_{\bar{A}}i} | L^{\bar{A}}_{\boldsymbol{g}'_{\bar{A}} i'} \rangle=\langle \phi^{\bar{A}}_i | \phi^{\bar{A}}_{i'} \rangle$ after imposing the additional constraint on the gauge charges in $\bar{A}$, such that the summation only involves sectors where $\boldsymbol{g}_{\bar{A}}$ and $\boldsymbol{g}'_{\bar{A}}$ can differ only at the boundary site.
For generic initial state with same fermionic configuration, we have $|z(0)|>0$.

\textbf{Initial entanglement asymmetry.}
We evaluate the EA at $t=0$. We consider an initial state with the same fermionic configuration in each sector of the form:
\begin{equation}\label{SMini}
  |\Psi_{\text{ini}}\rangle=\frac{1}{\sqrt{d}}\sum_{\boldsymbol{g}_A}|\Psi_{\boldsymbol{g}_A}\rangle
  =\frac{1}{\sqrt{d}}\sum_{\boldsymbol{g}_Ai}\beta_{i}|\phi_i\rangle|L_{\boldsymbol{g}_Ai}\rangle,
\end{equation}
where $\beta_i$ is independent of $\boldsymbol{g}$, and $d$ is the number of sectors in the summation. Here, we take the gauge charges in $\bar{A}$ as zeros $\boldsymbol{g}_{\bar{A}}=\boldsymbol{0}_{\bar{A}}$.
The reduced density matrix is calculated as
\begin{equation}
\begin{aligned}
\rho_{A}&=
  \sum_{\boldsymbol{g}_{A},\boldsymbol{g}_{A}^{\prime}}\sum_{ii^\prime}
  \frac{\beta_i\beta_i^*}{d}\langle\phi^{\bar{A}}_{i}|
  \phi^{\bar{A}}_{i^\prime}\rangle\langle L^{\bar{A}}_{\boldsymbol{0}_{\bar{A}}i}|L^{\bar{A}}_{\boldsymbol{0}_{\bar{A}}i^{\prime}}\rangle|\phi^{A}_{i}
  \rangle\langle\phi^{A}_{i^\prime}|
   \otimes|L^{A}_{\boldsymbol{g}_{A}i}\rangle\langle L^{A}_{\boldsymbol{g}_{A}^{\prime}i^{\prime}}|\\
&=\sum_{\boldsymbol{g}_{A},\boldsymbol{g}_{A}^{\prime}}\sum_{ii^\prime}
  \frac{\beta_i\beta_i^*}{d}\langle\phi^{\bar{A}}_{i}|
  \phi^{\bar{A}}_{i^\prime}\rangle|\phi^{A}_{i}\rangle\langle\phi^{A}_{i^\prime}|
   \otimes|L^{A}_{\boldsymbol{g}_Ai}\rangle\langle L^{A}_{\boldsymbol{g}_A^{\prime}i^{\prime}}|\\
&=\sum_{ii^\prime}
  \frac{\beta_i\beta_i^*}{d}\langle\phi^{\bar{A}}_{i}|
  \phi^{\bar{A}}_{i^\prime}\rangle|\phi^{A}_{i}\rangle\langle\phi^{A}_{i^\prime}|
   \otimes\mathbf{1}_{d\times d}\\
&=\mathrm{Tr}_{\bar{A}}(\rho^f)\otimes\mathbf{1}_{d\times d}
\end{aligned}
\end{equation}
In the second equation, we have used the fact that the inner product
$\langle \phi^{\bar{A}}_i | \phi^{\bar{A}}_{i'} \rangle \langle L^{\bar{A}}_{\boldsymbol{0}_{\bar{A}} i} | L^{\bar{A}}_{\boldsymbol{0}_{\bar{A}} i'} \rangle$
is equal to $\langle \phi^{\bar{A}}_i | \phi^{\bar{A}}_{i'} \rangle$. Here, $\mathbf{1}_{d\times d}$ denotes the $d\times d$ matrix with all entries equal to unity. $\rho^f$ is the fermionic part of initial state. Consequently, for a fixed fermionic configuration, the reduced density matrix depends only on the sector number $d$.

\section*{Appendix II: Alternative metrics for quantifying state separation}
There exist several reasonable quantities that can be used to characterize the separation between two quantum states. In the main text, we employed the entanglement asymmetry to quantify the degree of symmetry breaking. Here, we demonstrate that alternative metrics exhibit the same relaxation behavior.

The first metric we examine is the trace distance, which has been used experimentally to probe both the strong and inverse quantum Mpemba effects in open trapped-ion systems:
\begin{equation}
  d_{\mathrm{Tr}}(t)
  =\frac{1}{2}\mathrm{Tr}\sqrt{[\rho_A(t)-\rho_{A,G}(t)][\rho_A(t)-\rho_{A,G}(t)]^{\dag}}.
\end{equation}

The second metric we consider is the R\'{e}nyi-2 entanglement asymmetry, which can be experimentally accessed via randomized measurements:
\begin{equation}
  \Delta S_A^{(2)}(t)
  =S^{(2)}[\rho_{A,G}(t)]-S^{(2)}[\rho_A(t)],
\end{equation}
where \(S^{(2)}[\rho]=-\,\log\mathrm{Tr}\,\rho^2\) denotes the R\'{e}nyi-2 entanglement entropy.

The numerical results are shown in Fig.~\ref{SMfig1} and Fig.~\ref{SMfig2}. We find that these alternative metrics reproduce the same relaxation behavior as the entanglement asymmetry, although their absolute values differ.

\begin{figure}[b]
  \centering
  \includegraphics[width=11cm]{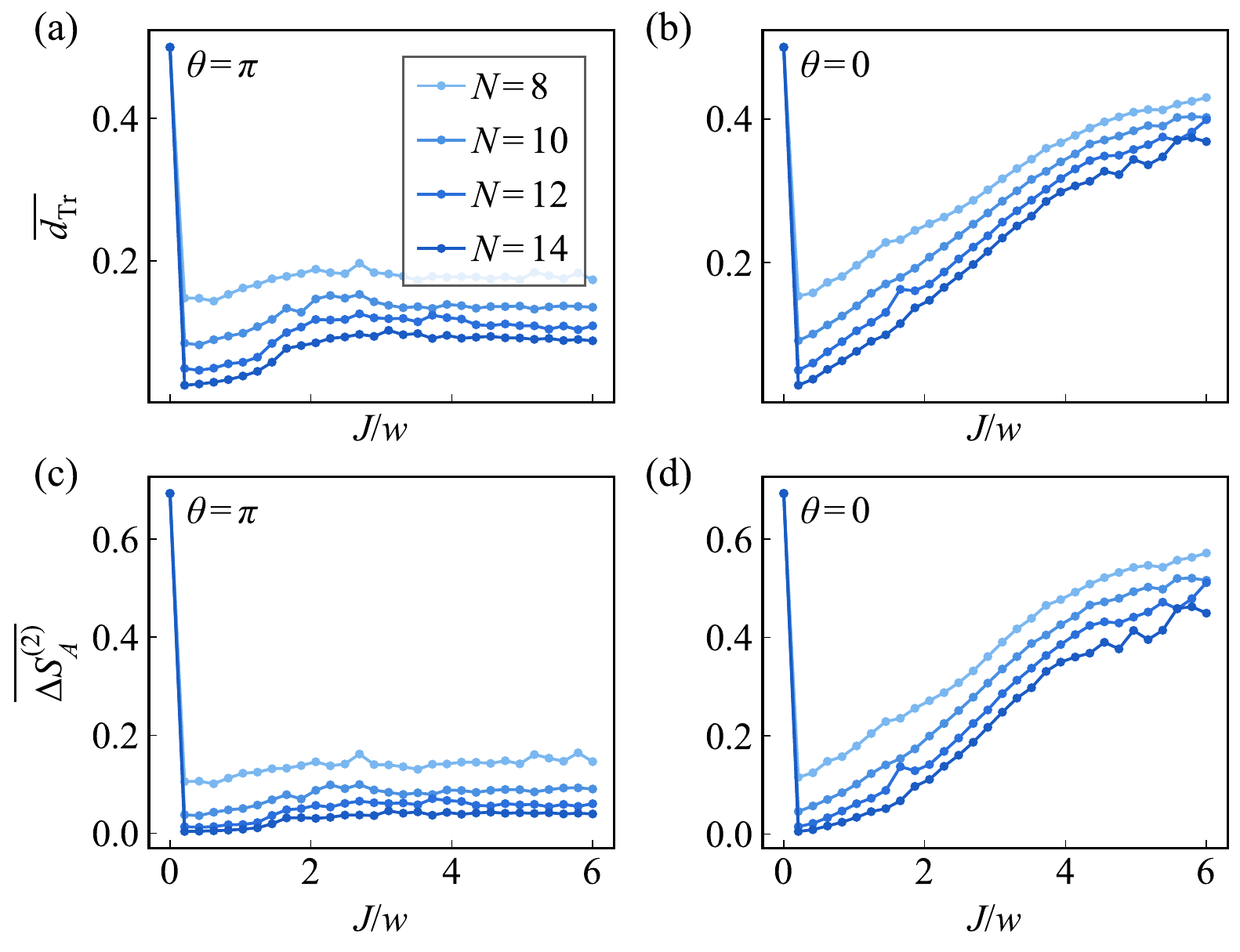}
  \caption{
  Long-time equilibrium values of the trace distance (a–b) and the R\'{e}nyi-2 entanglement asymmetry (c–d), averaged over \(wt=10^3\!-\!10^4\), as a function of the coupling strength \(J\) at topological angles \(\theta=\pi\) and \(0\) for different system sizes \(N\).
  }
  \label{SMfig1}
\end{figure}

\begin{figure}[t]
  \centering
  \includegraphics[width=11cm]{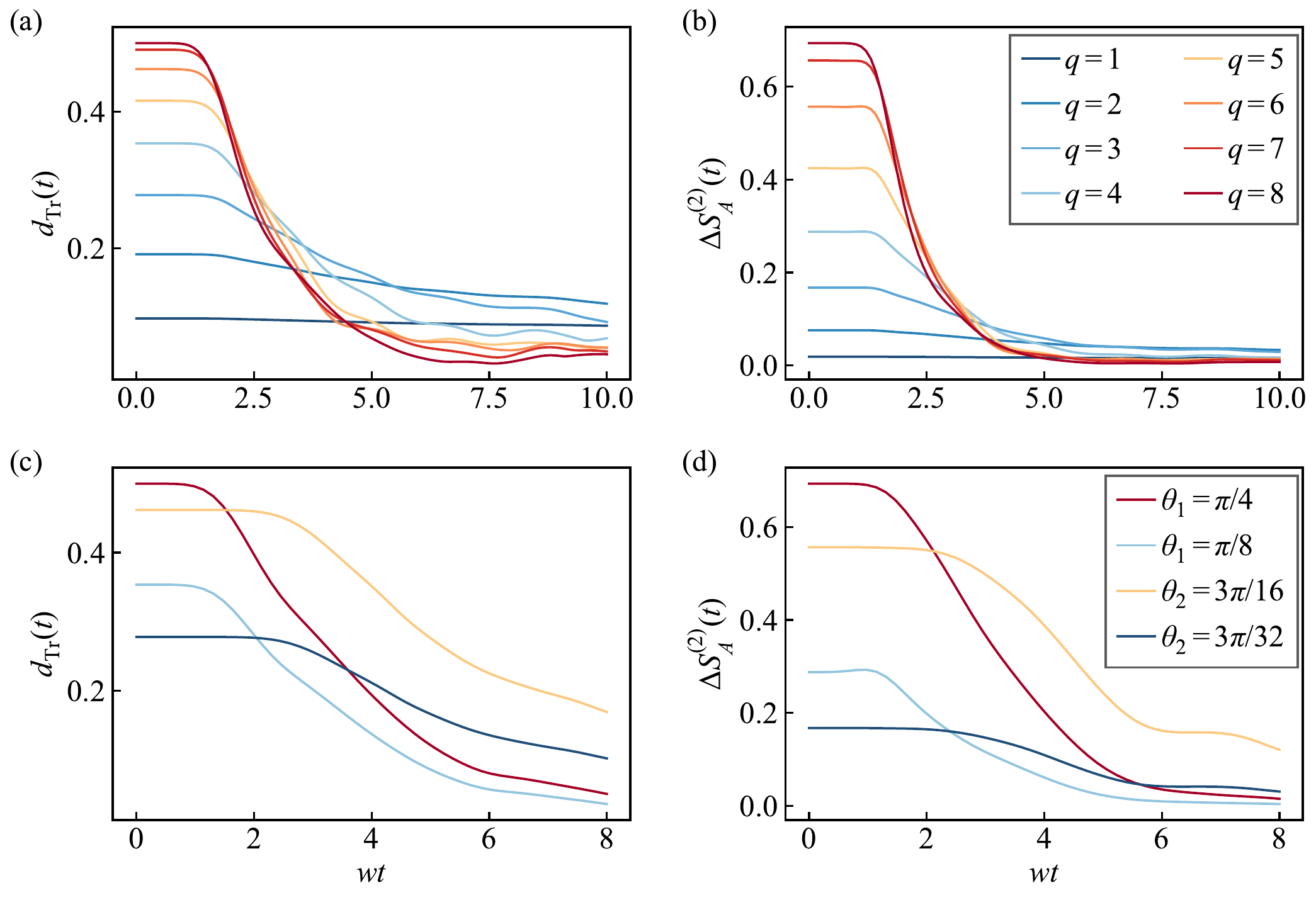}
  \caption{
  Dynamics of the trace distance in panels (a) and (c) and the R'enyi-2 entanglement asymmetry in panels (b) and (d).
  (a),(b) Subsystem dynamics of the lattice Schwinger model.
  (c),(d) Subsystem dynamics of the quantum link model.
  }
  \label{SMfig2}
\end{figure}

\end{widetext}
\end{document}